# Computation of homogenized stiffnesses of thin corrugated plates

Kolpakov A.A.[1)], Kolpakov A.G.[2)]

[1)] Institut de Mathématiques, Université de Neuchatel, Rue Emile-Argand 11, 2000 Neuchatel, Switzerland
e-mail: kolpakov.alexander@gmail.com

[2)] SysAn, A. Nevskogo Str., bld. 12a, suite 34, Novosibirsk, 630075, Russia
e-mail: algk@ngs.ru
*corresponding author*

**Abstract**. The homogenized stiffnesses of the corrugated plate are calculated by solving the periodicity cell problem of the homogenization theory by using two-steps dimension reduction procedure. The three-dimensional cell problem first reduces to a two-dimensional problem on the plate cross-sections. Then, provided that the plate is a thin, two-dimensional problem, it reduces to a one-dimensional problem, similar to the problem of curvilinear beam bending. Using the specifics of the arising one-dimensional problem, we construct its solution, depending on several constants.

As a result, the solution of the cell problem and the calculation of the homogenized stiffnesses are reduced to solving algebraic equations. It is found that essential calculations are necessary only for calculating the in-plane stiffnesses. The remaining effective stiffnesses are easily calculated by numerically calculating the integrals of known functions. Exact formulas are obtained for some effective stiffensses.

The paper is initialized by the observation that since the pioneering work of the 1920's until now numerous research publications have appeared that contain «new» formulas for computation of the homogenized stiffnesses of a corrugated plate. Those formulas differ one from another.

The fact that such a practically important problem known for about a century has not yet received a complete solution, arouses much interest and brings and begs for a definite answer. Its solution may be obtained on the basis of accurate and consistent application of elasticity theory and homogenization theory, which we provide in the present paper.

**Introduction**. The problem of computation of the homogenized (also referred to effective, averaged, macroscopic) stiffnesses of corrugated plates was first studied, apparently, in [1], and it keeps attracting attention to this day [2, 3, 4, 5, 6, 7, 8, 9, 10]. The recent progress in solving this problem is related to the development of homogenization theory [11, 12]. Homogenization theory on its own, although it has provided a theoretical basis for effective stiffnesses, could not provide any computational formulae: it reduced all computation to solving periodicity cell problems [11, 12]. A periodicity cell problem for a corrugate plate is a problem from elasticity theory, and thus can be tackled with some specific methods of the latter [8]. In the case of thin plates (which is a frequent property of corrugated plates), there have been methods developed based in the theory of shells and curvilinear beams. A survey of related results can be found in [6], which is also one of the most recent contribution to the subject, see also references in [13].

All the approaches known so far to the problem of computation of homogenized characteristics for thin-walled structures deal with a model whose dimension is already reduced: in the case of corrugated shells it's a two-dimensional model [5, 6, 14]. In the present work, we shall perform our computation of effective stiffnesses in two steps. First, we shall reduce a three-dimensional periodicity cell problem to a problem in the cross-section of the corrugated shell, as suggested in [8, 15]. This gives rise to a two-dimensional elasticity problem and a problem of anti-planar deformation in the cross-section of the corrugated plate, i.e. in a thin two-dimensional domain.

Second, we reduce the above problems to one-dimensional deformation and thermos-conductivity problems for a curvilinear beam: here an "approximation method" is used as described in [16]. Apparently, such a two-step dimension reduction is quite close to the method of asymptotic approximation for three-dimensional problems in thin domains [11, 12].

The paper pays significant attention to the mathematical justification of the developed procedure of computation for the homogenized characteristics of corrugated plate. At the same time, the authors opine that the article is important primarily for mechanics. The reason for it is as follows. The formulas obtained by previous authors often do not coincide with each other for either one or several stiffnesses, see the review in [6].

Such a discrepancy cannot be considered as a progress in the analysis of the problem, because both the original problem of elasticity theory for a corrugated plate and the homogenization procedure for it have unique solutions. In this case, various formulas can arise only through the use of approximations that are significantly different from each other or due to methodological differences in solving the problem. While any methods that preserve the level of rigor of elasticity theory and homogenization theory are used, the formulas obtained must coincide.

Therefore, it seems that developing a mathematically rigorous and clear method for solving this problem is quite an actual need in the mechanics of plates of inhomogeneous microstructure.

The authors suggest a solution and a computational method. The advantage of this method is that it specifies the curvature $\rho(s)$ of the corrugation with natural parameterization ($s$ means the path length along the corrugation) as the only control function for the problem. This simplification transforms the original problem into a system of linear ordinary differential equations with constant coefficients by substitution $d\varphi = \rho(s)ds$. However, the main idea is the use of the curvature $\rho(s)$ and the angle $\varphi(s)$ as principal papameters of the problem, and not the use of the natural parameterization.

Such a simplification seems important, because the solution below does not leave any hope for the existence of more or less simple explicit formulas for computation the in-plane stiffnesses for arbitrary corrugation profiles. This is due to the need of "gluing" of solutions at the points of curvature's change: the curvature of a "macroscopically flat" plate necessarily changes its sign. One can solve the corresponding system of algebraic "gluing equations" explicitly. However, the resulting formulas are so cumbersome that they are unlikely to be of interest to both practical engineers and mathematicians (in mathematics, other methods are used to analyze such situations, for example, methods of optimal control for continuous systems [17], and not direct analysis of the formulas).

The calculation of the homogenized in-plane stiffnesses can be effectively implemented numerically for arbitrary corrugation profiles, but this represents a topic for a separate paper.

The authors present an explicit formula for computation of the homogenized stiffness in the direction orthogonal to the corrugations profile for one particular type of corrugation. The fast that this formula acyually coincides with one in the existing literature, and that it is confirmed by finite element calculations, indicates only the absence of errors in our method, but does not imply the possibility of obtaining explicit formulas in the general case.

For stiffnesses, except the homogenized in-plane stiffnesses, explicit formulas or expressions in the form of integrals of one-variable functions may be obtained.

Let us consider a periodic corrugated plate which is represented by the three-dimensional domain depicted in Fig.1. There the principle direction $L$ of corrugation is parallel to the $Ox_1$ axis, and $C$ denotes the cross-section of the plate. Let $P$ denote the periodicity cell of the plate as a three-dimensional body (highlighted in Fig.1). The chosen periodicity cell can be represented as $P = [0,1] \times P_0$, where $P_0$ is the periodicity cell of plate's section $C$.

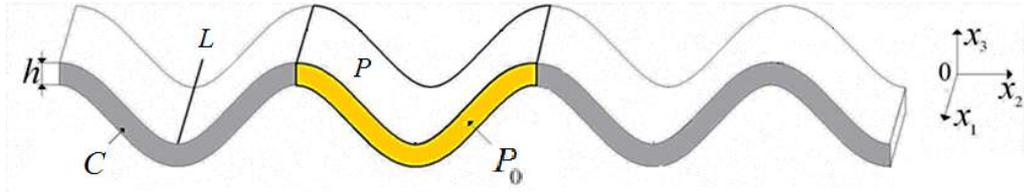

**Fig.1** Corrugated plate and its cross-section

If the characteristic dimensions of the corrugated plate (the length of the corrugation wave and its height, see Fig.1) are small then we can associate the given corrugated plate and a flat plate with the governing equations

$$N_{\alpha\beta} = D^1_{\alpha\beta\gamma\delta}\xi_{\gamma\delta} + D^0_{\alpha\beta\gamma\delta}e_{\gamma\delta}, \quad M_{\alpha\beta} = D^2_{\alpha\beta\gamma\delta}\xi_{\gamma\delta} + D^1_{\alpha\beta\gamma\delta}e_{\gamma\delta}, \tag{1}$$

where $e_{\gamma\delta}$ and $\xi_{\gamma\delta}$ are the homogenized (macroscopic) deformations in the plane of the homogenized (frat) plate and its curvature/torsion, $M_{\alpha\beta}$ and $N_{\alpha\beta}$ are the moments and forces, while the following are the stiffnesses: $D^0_{\alpha\beta\gamma\delta}$ - in the plane of the plate, $D^2_{\alpha\beta\gamma\delta}$ - bending/torsion and $D^1_{\alpha\beta\gamma\delta}$ - asymmetric.

Since the domain in our case is cylindrical with axis $Ox_1$, all the respective periodicity cell problems about the shape of the corrugation wave (see Fig.2, left) and its deformations under the main types of global deformations are reduced to two-dimensional problems in the cross-section on the plate (see Fig.2, right).

The periodicity cell of our plate is a thin curvilinear beam of thickness $h$, whose medial line is given by the equation $\mathbf{x} = \mathbf{x}(s) = (x_2(s), x_3(s))$, c.f. Fig.2, right. Let us assume that the medial line is a smooth curve (i.e. without any "corner points"), which keeps symmetric with respect to the corrugation. Let $L$ be the half-length of the corrugation wave, while the length of its projection on the $Oy_2$ axis be $P$, see Fig.2.

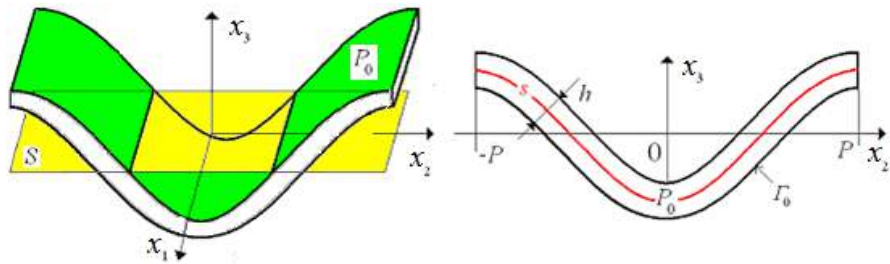

**Fig.2** 3-D periodicity cell and its 2-D cross-section

The medial line of the corrugation has a natural parameter $s$, which is the length measured along it, and a natural coordinate system with unit vectors (respective, a tangent one and a normal one) $\{\boldsymbol{\tau}, \mathbf{n}\}$, such that $\boldsymbol{\tau} = \dfrac{d\mathbf{x}}{ds}(s)$. The system $\{\boldsymbol{\tau}, \mathbf{n}\}$ is orthonormal. Let $n$ be the coordinate in the direction of $\mathbf{n}$, and let $u(s)$ and $w(s)$ be the tangential and normal displacements, respectively.

**1. The effective tension stiffness and the effective bending stiffness in the direction across the corrugation profile.** As we shall see later, most computational effort is invested into the calculation of effective tension stiffnesses across the corrugation profile (i.e. in the plane of the section depicted in Fig.2) Thus, a good deal of our work concerns the tension of the corrugation waves in the $Ox_2$ direction.

**1.1. Periodicity problems in two-dimensional elasticity theory.** There are two cases in which two-dimensional periodicity problems arise in the elasticity theory for corrugated plates: a periodicity problem for the tension along the $Ox_2$ axis (for computing the effective in-plane stiffness $D_{2222}^0$) and a one for bending in the plane orthogonal to the plate's section (for computing its effective bending stiffness $D_{2222}^2$). In other cases only anti-plane deformation problems arise, for which we arrive at a single equation. Another way would be using the universal relations between plate's homogenized stiffnesses and avoid any periodicity problems altogether.

Any of the above mentioned periodicity problems are then reduced to a two-dimensional elasticity problem on the plate's section. In the present notation, the periodicity cell problem takes the form

$$\begin{cases} (a_{ijkl} N_{k,l}^\nu + a_{ij22}(-1)^\nu x_3^\nu)_{,j} = 0 \quad \text{in } P_0, & (2.1) \\ (a_{ijkl} N_{k,l}^\nu + a_{ij22}(-1)^\nu x_3^\nu) n_j = 0 \quad \text{on } \Gamma_0, & (2.2) \\ \mathbf{N}^\nu \text{ periodic in } x_2 \text{ on } [-P, P]. & (2.3) \end{cases}$$

$\mathbf{N}^\nu = (N_2^\nu, N_3^\nu)(x_2, x_3)$ is the two-dimensional displacement vector in the $Ox_2 x_3$ plane, the index $\nu$ takes 0 and 1: $\nu = 0$ corresponds to the tension along $Ox_2$, $\nu = 1$ corresponds to the bending in the $Ox_2 x_3$ plane, in the direction orthogonal to the plate's section (across the corrugation waves).

It is known that the respective displacements exist if the deformations are given by $e_{ij} = (-1)^\nu x_3^\nu \delta_{i2} \delta_{j2}$, c.f. [18]. Then one can introduce such functions $\mathbf{M}^\nu = (M_2^\nu, M_3^\nu)(x_2, x_3)$ that (2.1) and (2.2) take the following form (c.f. [8])

$$\begin{cases} (a_{\gamma\delta\iota\kappa} M_{\iota,\kappa}^\nu (x_2, x_3))_{,\delta} = 0 \quad \text{in } P_0, & (3.1) \\ a_{\gamma\delta\iota\kappa} M_{\iota,\kappa}^\nu (x_2, x_3) n_\delta = 0 \quad \text{on } \Gamma_0. & (3.2) \end{cases}$$

where

$$M_{\iota,\kappa}^\nu = N_{\iota,\kappa}^\nu + (-1)^\nu \delta_{\iota\kappa 22} x_3^\nu$$

If the corrugation is symmetric with respect to the medial line of the plate (across the corrugation waves), then the periodicity condition (2.3) is replaced by

$$M_2^\nu(\pm P, x_3) = \pm(-1)^\nu P x_3^\nu \tag{3.3}$$

The stiffness $D_{2222}^{\nu+\mu}$ can be computed as [11, 12]

$$D_{2222}^{\nu+\mu} = \frac{1}{2P} \int_{P_0} (a_{2222}(-1)^\nu x_3^\nu + a_{22\iota\kappa} N_{\iota,\kappa}^\nu)(-1)^\mu x_3^\mu dx_2 dx_3 = \frac{1}{2P} \int_{P_0} a_{22\iota\kappa} M_{\iota,\kappa}^\nu (-1)^\mu x_3^\mu dx_2 dx_3 \tag{4}$$

From (2) we obtain the equality

$$\int_{P_0} (a_{\gamma\delta\iota\kappa} N_{\iota,\kappa}^\nu N_{\gamma,\delta}^\nu + (-1)^\nu a_{\gamma\delta 22}(x_2, x_3) x_3^\nu N_{\gamma,\delta}^\nu) dx_2 dx_3 = 0 \tag{5}$$

and thus

$$a_{\gamma\delta\iota\kappa} N_{\iota,\kappa}^\nu N_{\gamma,\delta}^\nu = a_{\gamma\delta\iota\kappa}(M_{\iota,\kappa}^\nu - (-1)^\nu \delta_{\iota\kappa 22} x_3^\nu)(M_{\gamma,\delta}^\nu - (-1)^\nu \delta_{\gamma\delta 22} x_3^\nu) =$$
$$= a_{\gamma\delta\iota\kappa} M_{\iota,\kappa}^\nu M_{\gamma,\delta}^\nu - 2(-1)^\nu x_3^\nu a_{\gamma\delta 22} N_{\gamma,\delta}^\nu - a_{2222} x_3^{2\nu}.$$

By using (5), we obtain further

$$\int_{P_0} (a_{\gamma\delta\iota\kappa} M_{\iota,\kappa}^\nu M_{\gamma,\delta}^\nu - 2(-1)^\nu x_3^\nu a_{\gamma\delta 22} N_{\gamma,\delta}^\nu - a_{2222} x_3^{2\nu}) dx_2 dx_3 = -\int_{P_0} (-1)^\nu a_{\gamma\delta 22} x_3^\nu N_{\gamma,\delta}^\nu dx_2 dx_3$$

This yields that

$$\int_{P_0} a_{\gamma\delta\iota\kappa} M^\nu_{\iota,\kappa} M^\nu_{\gamma,\delta} dx_2 dx_3 = \int_{P_0} (a_{2222} x_3^{2\nu} + (-1)^\nu a_{\gamma\delta 22} x_3^\nu N^\nu_{\gamma,\delta}) dx_2 dx_3$$

and the stiffnesses $D^{\nu+\mu}_{2222}$ in (4) can be computed via the elastic energy

$$D^\nu_{2222} = \frac{1}{2P} \int_{P_0} a_{\gamma\delta\iota\kappa} M^\nu_{\iota,\kappa} M^\nu_{\gamma,\delta} dx_2 dx_3 \tag{6}$$

The boundary condition (3.3) corresponds to the unit relative displacement or to the unit torsion of the plate's ends. (see Fig.2). If the relative displacement or torsion equals $A$ instead, then (3.3) becomes $M^\nu_2(\pm P, x_3) = \pm(-1)^\nu APx_3^\nu$, and (6) subsequently turns into

$$\frac{1}{2} D^\nu_{2222} A^2 \cdot 2P = \frac{1}{2} \int_{P_0} a_{\gamma\delta\iota\kappa} M^\nu_{\iota,\kappa} M^\nu_{\gamma,\delta} dx_2 dx_3 \tag{7}$$

Here $A$ is actually the homogenized deformation $e_{22}$ of the homogenized plate if $\nu = 0$ and the curvature $\xi_{22}$ if $\nu = 1$. The multiplier $2P$ is the length of the projection of the corrugation wave onto the $Ox_2$ axis. This means that $\frac{1}{2} D^\nu_{2222} A^2 \cdot 2P$ is the energy of a single corrugation wave of the homogenized plate. Now formula (7) expresses the fact that the original corrugated plate and the homogenized flat plate have equal energies, which is one the main principles in homogenization theory as applied to the inhomogeneous elastic materials and structures [19]. We deduced it in our specific case to make use of later in the sequel.

**1.2. Equilibrium equations for curvilinear beam corresponding to a periodicity problem for a corrugation wave.** For periodicity problem such as (1) with $\nu = 0$ and $\nu = 1$ the equations (3.1) and (3.2) coincide except for their respective boundary conditions. Namely, we shall use the boundary conditions (3.3), which correspond to the tension of the periodicity cell along the $Ox_2$ axis (for $D^0_{2222}$) and bending in the plane transversal to the corrugation cross-section (in order to compute $D^2_{2222}$).

Let us derive the equilibrium conditions. Since the profile of the corrugation is given by some functions ($x(s), y(s)$) its curvature $\rho(s)$ is a known function. The curvature radius of the initial medial section $R(s) = 1/\rho(s)$ is much bigger than the plate's thickness, so that the thin shell theory applies [20]. We shall use its methods in order to obtain a boundary problem for a curvilinear beam (which corresponds to a two-dimensional periodicity problem). In this case deformation in the fiber parallel to the beam's middle (neutral) surface

$$e = e_0 - n\varpi, \tag{8}$$

where its middle (neutral) surface deformation is

$$e_0 = u'(s) - w(s)\rho(s), \tag{9}$$

and the change in the curvature equals

$$\varpi = [w'(s) + u(s)\rho(s)]'. \tag{10}$$

The quantity

$$y = w'(s) + u(s)\rho(s) \tag{11}$$

is the rotation angle of the tangent vector $\tau$ (or, equivalently, of the normal $n$), see Fig. 1.

The two-dimensional periodicity cell problem for the corrugation wave corresponds to the planar stressed state - displacement orthogonally to its cross-section is zero, see [8, 15]. The corresponding potential energy of the deformation (8) can be expressed as

$$J(u,w) = \frac{1}{2}\int_{-h/2}^{h/2}\int_0^L \frac{E}{1-\nu^2} e^2 ds = \frac{1}{2}\int_{-h/2}^{h/2}\int_0^L \frac{E}{1-\nu^2}[u'-w\rho-(w'+u\rho)'n]^2 dsdn \qquad (12)$$

The equilibrium conditions follow from the fact that the variation of (12) is vanishing:

$$\delta J = \int_{-h/2}^{h/2}\int_0^L \frac{E}{1-\nu^2}[u'(s)-w(s)\rho(s)-(w'(s)+u(s)\rho(s))'n]\times$$
$$\times[\delta u'(s)-\delta w(s)\rho(s)-(\delta w'(s)+\delta u(s)\rho(s))'y]dsdn = 0 \qquad (13)$$

Integration (13) with respect to $n$ yields

$$\frac{Eh}{1-\nu^2}\int_0^L [u'(s)-w(s)\rho(s)][\delta u'(s)-\delta w(s)\rho(s)]ds + \qquad (14)$$
$$+\frac{Eh^3}{12(1-\nu^2)}\int_0^L [w'(s)+u(s)\rho(s)]'\times[\delta w'(s)+\delta u(s)\rho(s)]' ds = 0.$$

Integrating (14) further by parts, we obtain

$$-\frac{Eh}{1-\nu^2}[u'(s)-w(s)\rho(s)]' - \frac{Eh^3}{12(1-\nu^2)}[(w'(s)+u(s)\rho(s))]''\rho(s) = 0, \qquad (15)$$

$$-\frac{Eh}{1-\nu^2}[u'(s)-w(s)\rho(s)]\rho(s) + \frac{Eh^3}{12(1-\nu^2)}[(w'(s)+u(s)\rho(s))]''' = 0.$$

The above equations (15) are the one-dimensional equations for the approximate problem to the initial two-dimensional problem (3).

Let us transform (16) into a system of first order equations. Let's first introduce the notation

$$x(s) = u'(s) - w(s)\rho(s), \quad y(s) = w'(s) + u(s)\rho(s). \qquad (16)$$

Then (15) takes the form

$$x' + Hy''\rho(s) = 0, \quad -x\rho(s) + Hy''' = 0, \qquad (17)$$

with $H = h^2/12$. Let's also put

$$\xi(s) = y'(s), \quad \eta(s) = \xi'(s). \qquad (18)$$

With notation (18) the equations in (17) become a first order system of six equations about the six functions $u, w, y, \xi, x, \eta$ (which we group in pairs for convenience).

$$\begin{cases} u' = w\rho(s) + x, \\ w' = -u\rho(s) + y, \end{cases} \qquad (19.1)$$

$$\begin{cases} y' = \xi, \\ \xi' = \eta, \end{cases}$$

$$\begin{cases} x' = -H\eta\rho(s), \\ \eta' = H^{-1}x\rho(s). \end{cases} \qquad (19.2)$$

$$(19.3)$$

The elastic energy (6) for a half-wave is

$$J(x,\xi) = \frac{Eh}{1-\nu^2}\int_0^L x^2(s)ds + \frac{Eh^3}{12(1-\nu^2)}\int_0^L \xi^2(s)ds. \qquad (20)$$

The quantities in (14) and (20) have the following meaning (in addition to what is already explained above): $N = \frac{Eh}{1-\nu^2}x$ is the resultant axial force, $x$ is the axial deformation $e_0$, $Q = \frac{Eh^3}{12(1-\nu^2)}\eta$ is the sheer force, and $\xi$ is the change of the beam's axial curvature.

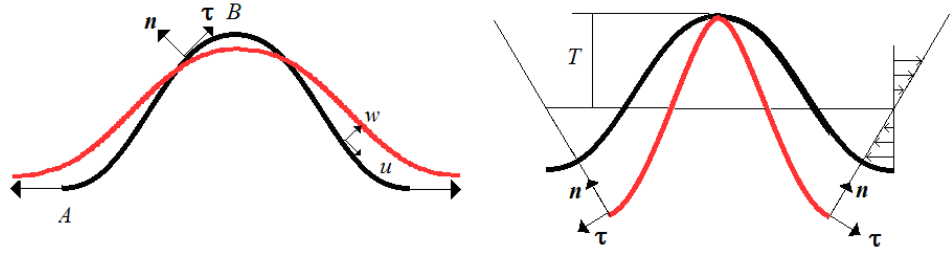

**Fig.3** Tension and bending of the periodicity cell

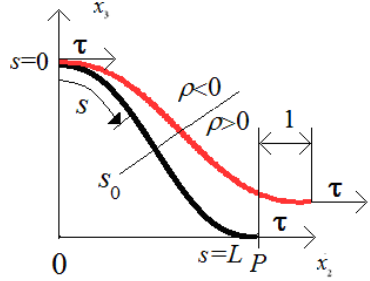

**Fig.4** Corrugation half-wave

**1.3. Computing the tensile stiffness $D^0_{2222}$.** As already noted, the equilibrium conditions (14) are the same for the bending and tension of the corrugation wave, the only difference being the boundary conditions. Let us write down the boundary conditions for computing the tension of the corrugation wave depicted in Fig.3 on the left. The points A and B of the corrugation can freely move in the upward and downward directions, and the sheer force $Q = \eta$ at these points is equal to zero.

| | | | |
|---|---|---|---|
| $u(0) = 0$ | (21.1) | $u(L) = 1$ | (21.5) |
| $w(0) = 0$ | (21.2) | $y(L) = 0$ | (21.6) |
| $Q(0) = 0 \leftrightarrow \eta(0) = 0$ | (21.3) | | |
| $y(0) = 0$ | (21.4) | | |

The above condition (21.2) can be achieved by moving the half-wave as a solid, which does not influence the equality $Q(0) = 0$.

The system (14), and thus the mechanical properties of the corrugated plate, depend only one a single functional parameter: the curvature $\rho(s)$ of the corrugation profile. Here, the curvature $\rho(s)$ is a non-constant function, since $\rho(s) = const$ defines a circle, and thus precludes us from creating a plate.

The angle between the tangent vector $\tau$ to the medial line of the corrugation and the axis $Ox_2$ equals $\varphi(s) = \int_0^s \rho(t)dt$. We shall consider only smooth corrugation plates without creases (where $\rho(s) = \infty$) or flats (where $\rho(s) = 0$ on some intervals). For the cases where such things as creases and flats are allowed, c.f. [21]. In our case, since the corrugation wave is symmetric and smooth, we have that $\varphi(0) = 0$. We shall use $\varphi(s)$ as a new variable: however passing to it is possible only on segments where $\varphi(s)$ is monotonic, i.e. where $\rho(s)$ does not change its sign.

We shall consider the case where the corrugation wave is symmetric, and the corresponding half-wave covers two segments of monotonicity for $\varphi(s)$. Without loss of generality, we assume that $\rho < 0$ for $0 \leq s \leq s_0$ and $\rho > 0$ for $s_0 \leq s \leq L$, see Fig.4. A generic profile of such corrugation is shown in Fig.1. This case is very widespread in practice, and our method is applicable for a bigger number of monotonicity intervals.

**Solution to the problem (19), (21) for** $0 \leq s \leq s_0$ (i.e. while $\rho < 0$). The system of equations (19.3), once we put $d\varphi = \rho(s)ds$, becomes

$$\begin{cases} \dfrac{dx}{d\varphi} = -H\eta, \\ \dfrac{d\eta}{d\varphi} = H^{-1}x. \end{cases} \tag{22}$$

The characteristic polynomial of (22) equals $\lambda^2 + 1 = 0$, whose roots are $\lambda = \pm i$. Then the general solution to (22) has the form [22]

$$\begin{pmatrix} x \\ \eta \end{pmatrix} = C_1 \begin{pmatrix} H\cos\varphi(s) \\ \sin\varphi(s) \end{pmatrix} + C_2 \begin{pmatrix} H\sin\varphi(s) \\ -\cos\varphi(s) \end{pmatrix}.$$

Using the boundary condition (21.3) together with the fact that $\varphi(0) = 0$ leads to $\eta(0) = -C_2 = 0$, and thus

$$\begin{pmatrix} x \\ \eta \end{pmatrix} = C_1 \begin{pmatrix} H\cos\varphi(s) \\ \sin\varphi(s) \end{pmatrix}. \tag{23}$$

By using the equations (19.2), we obtain

$$\xi(s) = C_1 \int_0^s \eta(t)dt = C_1 \int_0^s \sin\varphi(t)dt + \xi_0 = C_1 k_{\sin}(s) + \xi_0, \tag{24}$$

$$y(s) = \int_0^s \xi(t)dt = C_1 \int_0^s k_{\sin}(t)dt + \xi_0 s + y_0 = C_1 K_{\sin}(s) + \xi_0 s + y_0.$$

In (24) we use the functions

$$k_{\sin}(s) = \int_0^s \sin\varphi(t)dt, \quad K_{\sin}(s) = \int_0^s k_{\sin}(t)dt. \tag{25}$$

Since (21.4) holds, and $K_{\sin}(0) = 0$, it follows that we have $y_0 = 0$ in (24).

Let us now consider (19.1). The corresponding homogeneous system, once we put $d\varphi = \rho(s)ds$, takes the form

$$\begin{cases} \dfrac{du}{d\varphi} = w, \\ \dfrac{dw}{d\varphi} = -u. \end{cases} \tag{26}$$

The characteristic polynomial of (26) equals $\lambda^2 + 1 = 0$, and has roots $\lambda = \pm i$. Thus, the general solution to (26) is

$$\begin{pmatrix} u \\ w \end{pmatrix} = B_1 \begin{pmatrix} \cos\varphi(s) \\ -\sin\varphi(s) \end{pmatrix} + B_2 \begin{pmatrix} \sin\varphi(s) \\ \cos\varphi(s) \end{pmatrix} \tag{27}$$

In order to obtain the general solution to (19.1) we need to find its particular solution. We shall seek it as the following ansatz:

$$V(s)\begin{pmatrix} \cos\varphi(s) \\ -\sin\varphi(s) \end{pmatrix} + W(s)\begin{pmatrix} \sin\varphi(s) \\ \cos\varphi(s) \end{pmatrix} \tag{28}$$

By substituting (28) into (19.1) and using $x(s) = C_1 H\cos\varphi(s)$, $y(s) = C_1 K_{\sin}(s) + \xi_0 s$ we obtain

$$V'(s)\begin{pmatrix} \cos\varphi(s) \\ -\sin\varphi(s) \end{pmatrix} + W'(s)\begin{pmatrix} \sin\varphi(s) \\ \cos\varphi(s) \end{pmatrix} = \begin{pmatrix} C_1 H\cos\varphi(s) \\ C_1 K_{\sin}(s) + \xi_0 s \end{pmatrix}. \tag{29}$$

By solving (29) we get
$$V'(s) = C_1 H \cos^2 \varphi(s) - C_1 K_{\sin}(s) \sin \varphi(s) - \xi_0 s \sin \varphi(s),$$
$$W'(s) = C_1 H \cos \varphi(s) \sin \varphi(s) + C_1 K_{\sin}(s) \cos \varphi(s) + \xi_0 s \cos \varphi(s).$$
(30)

Finally, it follows from (30) that
$$V(s) = \int_0^s [C_1 H \cos^2 \varphi(t) - C_1 K_{\sin}(t) \sin \varphi(t) + \xi_0 t \sin \varphi(t)] dt = C_1 v_0(s) - \xi_0 v_2(s),$$
(31)
$$W(s) = \int_0^s [C_1 H \cos \varphi(t) \sin \varphi(t) + C_1 K_{\sin}(t) \cos \varphi(t) + \xi_0 t \cos \varphi(t)] dt = C_1 w_0(s) + \xi_0 w_2(s)$$

where
$$v_0(s) = \int_0^s [H \cos^2 \varphi(t) - K_{\sin}(t) \sin \varphi(t)] dt, \quad v_2(s) = \int_0^s t \sin \varphi(t) dt,$$
(32)
$$w_0(s) = \int_0^s [H \cos \varphi(t) \sin \varphi(t) + K_{\sin}(t) \cos \varphi(t)] dt, \quad w_2(s) = \int_0^s t \cos \varphi(t) dt.$$

Thus (19.1) has a particular solution of the form
$$C_1 v_0(s) \begin{pmatrix} \cos \varphi(s) \\ -\sin \varphi(s) \end{pmatrix} - \xi_0 v_2(s) \begin{pmatrix} \cos \varphi(s) \\ -\sin \varphi(s) \end{pmatrix} + C_1 w_0(s) \begin{pmatrix} \sin \varphi(s) \\ \cos \varphi(s) \end{pmatrix} + \xi_0 w_2(s) \begin{pmatrix} \sin \varphi(s) \\ \cos \varphi(s) \end{pmatrix}.$$
(33)

Then (27) and (33) provide the general solution to (19.1), which is
$$\begin{pmatrix} u \\ w \end{pmatrix} = B_1 \begin{pmatrix} \cos \varphi(s) \\ -\sin \varphi(s) \end{pmatrix} + B_2 \begin{pmatrix} \sin \varphi(s) \\ \cos \varphi(s) \end{pmatrix} + C_1 \begin{pmatrix} U_1(s) \\ W_1(s) \end{pmatrix} + \xi_0 \begin{pmatrix} U_2(s) \\ W_2(s) \end{pmatrix},$$
(34)

where
$$\begin{pmatrix} U_1(s) \\ W_1(s) \end{pmatrix} = v_0(s) \begin{pmatrix} \cos \varphi(s) \\ -\sin \varphi(s) \end{pmatrix} + w_0(s) \begin{pmatrix} \sin \varphi(s) \\ \cos \varphi(s) \end{pmatrix},$$
(35)
$$\begin{pmatrix} U_2(s) \\ W_2(s) \end{pmatrix} = v_2(s) \begin{pmatrix} \cos \varphi(s) \\ -\sin \varphi(s) \end{pmatrix} + w_2(s) \begin{pmatrix} \sin \varphi(s) \\ \cos \varphi(s) \end{pmatrix}.$$

We shall use the boundary conditions (21.1), (21.2) and the fact that at $s = 0$ we have that the functions in (32) vanish: $\varphi(0) = 0$, $v_0(0) = 0$, $v_2(0) = 0$, $w_0(0) = 0$, and $w_2(0) = 0$. Then (34), (35), (21.1) and (21.2) yield
$$\begin{pmatrix} u \\ w \end{pmatrix}(0) = B_1 \begin{pmatrix} 1 \\ 0 \end{pmatrix} + B_2 \begin{pmatrix} 0 \\ 1 \end{pmatrix} = \begin{pmatrix} 0 \\ 0 \end{pmatrix},$$
which immediately implies that $B_1 = B_2 = 0$. Then
$$\begin{pmatrix} u \\ w \end{pmatrix}(s) = C_1 \begin{pmatrix} U_1(s) \\ W_1(s) \end{pmatrix} + \xi_0 \begin{pmatrix} U_2(s) \\ W_2(s) \end{pmatrix} =$$
(36)
$$= C_1 v_0(s) \begin{pmatrix} \cos \varphi(s) \\ -\sin \varphi(s) \end{pmatrix} + C_1 w_0(s) \begin{pmatrix} \sin \varphi(s) \\ \cos \varphi(s) \end{pmatrix} - \xi_0 v_2(s) \begin{pmatrix} \cos \varphi(s) \\ -\sin \varphi(s) \end{pmatrix} + \xi_0 w_2(s) \begin{pmatrix} \sin \varphi(s) \\ \cos \varphi(s) \end{pmatrix}.$$

The constants $C_1$ and $\xi_0$ in the above solution to the problem (14), (16) with $0 \leq s \leq s_0$ are yet to be determined.

**Solution to the problem (14), (16) for $s_0 \leq s \leq L$** (i.e. while $\rho > 0$). In this case we need to obtain a solution to (14) for $s_0 \leq s \leq L$ and make it merge at $s = s_0$ with the previously obtained one for $s < s_0$.

The solution for the case where $s_0 \leq s \leq L$ is obtained in complete analogy with the previous section and has a similar form. The angle between the tangent vector and the $Ox$ axis is equal to $\varphi(s) = \varphi_0 + \int_{s_0}^{s} \rho(t)dt$, where $\varphi_0 = \varphi(s_0)$. The main problem is merging the two obtained solutions at $s = s_0$ in such way that the resultant function be continuous.

The general solution to (19.3) for $s_0 \leq s \leq L$ has the form
$$\begin{pmatrix} x \\ \eta \end{pmatrix}(s) = C_1^+ \begin{pmatrix} H\cos\varphi(s) \\ \sin\varphi(s) \end{pmatrix} + C_2^+ \begin{pmatrix} H\sin\varphi(s) \\ -\cos\varphi(s) \end{pmatrix}. \tag{37}$$

The index «+» indicates that the corresponding constants belong to the case where $s > s_0$. The $k^+$-functions are defined analogously to those in (24) while substituting $\int_0^s dt$ with $\int_{s_0}^s dt$.

We shall equate the values of the functions $x$ and $\eta$ given by (37) with those of $x$ and $\eta$ given by (23) at $s = s_0$. Thus, we obtain
$$\begin{aligned} C_1^+ H\cos\varphi_0 + C_2^+ H\sin\varphi_0 &= C_1 H\cos\varphi_0, \\ C_1^+ \sin\varphi_0 - C_2^+ \cos\varphi_0 &= C_1 \sin\varphi_0. \end{aligned} \tag{38}$$

By solving (38) one gets $C_1^+ = C_1$, $C_2^+ = 0$, which means that $\eta(s)$ and $x(s)$ are given by (23) also for $s_0 \leq s \leq L$, and thus for all $0 \leq s \leq L$. By taking (23) into account, the solution to (19.2) for $s_0 \leq s \leq L$ is
$$\xi(s) = C_1 k_{\sin}^+(s) + \xi_0^+, \tag{39}$$
$$y(s) = C_1 K_{\sin}^+(s) + \xi_0^+ s + y_0^+$$

Merging the functions $\xi$, $y$ given by (39) with the functions $\xi$, $y$ given by (24), together with the fact that $y_0 = 0$ and $k_{\sin}^+(s_0) = 0$, $K_{\sin}^+(s_0) = 0$ yields
$$\begin{aligned} C_1 k_{\sin}(\varphi_0) + \xi_0 &= \xi_0^+. \\ C_1 K_{\sin}(\varphi_0) + \xi_0 s_0 &= \xi_0^+ s_0 + y_0^+. \end{aligned} \tag{40}$$

The functions $u$, $w$ for $s_0 \leq s \leq L$ can be found analogously to (34) and have the form
$$\begin{pmatrix} u \\ w \end{pmatrix} = B_1^+ \begin{pmatrix} \cos\varphi(s) \\ -\sin\varphi(s) \end{pmatrix} + B_2^+ \begin{pmatrix} \sin\varphi(s) \\ \cos\varphi(s) \end{pmatrix} + C_1 \begin{pmatrix} U_1^+(s) \\ W_1^+(s) \end{pmatrix} + \xi_0 \begin{pmatrix} U_2^+(s) \\ W_2^+(s) \end{pmatrix}.$$

Merging these functions $u$, $w$ with the functions $u$, $w$ given by (36) together with $U_1^+(s_0) = 0$, $U_2^+(s_0) = 0$, $V_1^+(s_0) = 0$, $V_2^+(s_0) = 0$ yields
$$\begin{aligned} B_1^+ \cos\varphi_0 + B_2^+ \sin\varphi_0 &= C_1 U_1(s_0) + \xi_0 W_1(s_0), \\ -B_1^+ \sin\varphi_0 + B_2^+ \cos\varphi_0 &= C_1 U_2(s_0) + \xi_0 W_2(s_0). \end{aligned} \tag{41}$$

Since $\cos\varphi(L) = 1$ and $\sin\varphi(L) = 0$, the boundary conditions (21.5), (21.6) yield
$$B_1^+ + C_1 U_1^+(L) + \xi_0 W_1^+(L) = 1, \tag{42.1}$$
$$C_1 K_{\sin}^+(L) + \xi_0^+ L + y_0^+ = 0. \tag{42.2}$$

We still need to determine the constants $B_1^+$, $B_2^+$, $\xi_0^+$, $y_0^+$ in the solution obtained for $s_0 \leq s \leq L$, as well as the constants $C_1$ and $\xi_0$ in the solution for $0 \leq s \leq s_0$. Thus, we have yet six undetermined constants. As well we have the six equations in (41), (40) and (29). From (41) we find

$$B_1^+ = (C_1 U_1(s_0) + \xi_0 V_1(s_0))\cos\varphi_0 - (C_1 U_2(s_0) + \xi_0 W_2(s_0))\sin\varphi_0, \quad (43)$$
$$B_2^+ = (C_1 U_1(s_0) + \xi_0 V_1^+(s_0))\sin\varphi_0 + (C_1 U_2(s_0) + \xi_0 W_2(s_0))\cos\varphi_0.$$

By substituting (43) into (42.1), we obtain
$$C_1 A + \xi_0 B = 1, \quad (44)$$
$$A = U_1(s_0)\cos\varphi_0 - U_2(s_0)\sin\varphi_0 + U_1^+(L),$$
$$B = V_1(s_0)\cos\varphi_0 - W_2(s_0)\sin\varphi_0 + W_1^+(L).$$

The functions $U_1, U_2(s_0), U_1^+(L), V_1, V_2, V_1^+$ are completely determined by (35).

So far we are left with the four equations in (40) and (42.2), namely
$$C_1 k_{\sin}(\varphi_0) + \xi_0 = \xi_0^+, \quad (45.1)$$
$$C_1 K_{\sin}(\varphi_0) + \xi_0 s_0 = \xi_0^+ s_0 + y_0^+, \quad (45.2)$$
$$C_1 K_{\sin}^+(L) + \xi_0^+ L + y_0^+ = 0. \quad (45.3)$$

The four equations from (44), (43) provide $C_1$, $\xi_0^+$, $y_0^+$ and $\xi_0$. From (45.3) we get
$$y_0^+ = C_1 K_{\sin}(\varphi_0) - s_0 C_1 k_{\sin}(\varphi_0) = C_1[K_{\sin}(\varphi_0) - s_0 k_{\sin}(\varphi_0)]. \quad (46)$$

From (45.3) and (46) it follows that
$$\xi_0^+ = -\frac{1}{L}(C_1 K_{\sin}^+(L) + y_0^+) = -\frac{C_1}{L}(K_{\sin}^+(L) + K_{\sin}(\varphi_0) - s_0 k_{\sin}(\varphi_0)). \quad (47)$$

From (45.2) and (47) moreover
$$\xi_0 = C_1[-\frac{1}{L}(K_{\sin}^+(L) + K_{\sin}(\varphi_0) - s_0 k_{\sin}(\varphi_0)) - k_{\sin}(\varphi_0)]. \quad (48)$$

From (44) and (48) we obtain
$$C_1 A + C_1[-\frac{1}{L}(K_{\sin}^+(L) + K_{\sin}(\varphi_0) - s_0 k_{\sin}(\varphi_0)) - k_{\sin}(\varphi_0)] = 1$$
and
$$C_1 = \frac{1}{A - \frac{1}{L}(K_{\sin}^+(L) + K_{\sin}(\varphi_0) - s_0 k_{\sin}(\varphi_0)) - k_{\sin}(\varphi_0) B}. \quad (49)$$

Thus the equalities (46)-(49) determine the constants $C_1$, $\xi_0^+$, $y_0^+$ and $\xi_0$.

The energy $J(x,\xi)$ is obtained by substituting our solution to the problem (14), (16) into (20). Since in this case $u(L) = 1$, the average deformation of the corrugation profile along the $Ox_2$ axis is $1/P$, where $P$ is the length of the corrugation half-wave's projection onto $Ox_2$. The effective tensile stiffness $D_{2222}^0$ of the corrugated plate along the $Ox_2$ axis is determined from the fact that the initial energy of the corrugated plate and that of the corresponding homogenized flat plate coincide [8, 15], i.e. $J(x,\xi) = \frac{1}{2} D_{2222}^0 \left(\frac{1}{P}\right)^2 P$ (the energy density multiplied by the length of the homogenized plate's specimen corresponding to the initial corrugated plate). We easily obtain
$$D_{2222}^0 = 2J(x,\xi)P. \quad (50)$$

Thus, the effective tensile stiffness $D_{2222}^0$ of the corrugated plate along the $Ox$ axis can be explicitly written by using the functions from (25) and (36). All the indicated functions are integrals of some known ones. The latter can be easily computed numerically, if not symbolically. The asymmetric stiffness equals

$$D_{2222}^{\nu+\mu} = \frac{1}{2P}\int_{P_0} a_{22\iota\kappa}M_{\iota,\kappa}^{\nu}(y_2,y_3)(-1)^{\mu}y_3^{\mu}dy_2dy_3 = \frac{1}{2P}\int_{P_0}\sigma_{22}^{\nu}(-1)^{\mu}y_3^{\mu}dy_2dy_3 =$$

$$= \frac{1}{P}\int_0^L\int_{-h/2}^{h/2}\sigma_{nn}\cos^2\varphi(s)(-1)^{\mu}(y_3+n\cos\varphi(s))^{\mu}dnds =$$

$$= -\frac{1}{P}\int_0^L\int_{-h/2}^{h/2}\frac{Eh}{(1-\nu^2)}(e_0-n\xi)\cos^2\varphi(s)(-1)^{\mu}(y_3+n\cos\varphi(s))^{\mu})dnds \quad (51)$$

Above, in (51), we use such $\nu$ and $\mu$ that $\nu+\mu=1$ and $\sigma_{22}^{\nu}=a_{22\iota\kappa}M_{\iota,\kappa}^{\nu}(y_2,y_3)$, while $\sigma_{nn}$ corresponds to $\sigma_{ij}^{\nu}$.

**2. Example**. Let us consider a corrugation profile which is composed from two quarters of circles having each radius $R$, as depicted in Fig.5 Each quarter-circle's length is $\pi R/2$, and the total half-wave's length is thus $\pi R$. For $0 \le s \le s_0 = \pi R/2$ the curvature equals $\rho = -1/R < 0$. We shall compute the effective tensile stiffness along the $Ox_2$ axis for this corrugation profile.

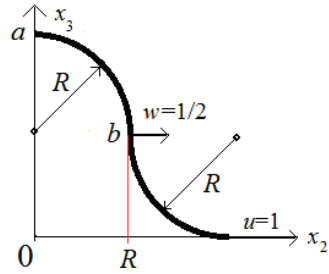

**Fig.5** Two quarter-circles constitute a corrugation profile

Because the corrugation half-wave is antisymmetric with respect to its midpoint $a$ in Fig.5, we shall consider a corrugation quarter-wave $ab$ instead. We have the boundary conditions (21.1) - (21.4) on its upper end $a$. On its right end $b$ we have the boundary conditions

$$\xi(\pi R/2) = 0 \leftrightarrow M(\pi R/2) = 0, \quad (52.1)$$
$$w(\pi R/2) = 1/2. \quad (52.2)$$

In particular, (52.1) means that the bending momentum at $b$ is zero, which is implied by the inner symmetry of the problem. The condition (52.2) is also implied by the symmetry, as it becomes obvious from Fig.5.

Thus, we need to find a solution only for $0 \le s \le s_0$ with $s_0 = \pi R/2$. The general solution that we obtained above for $0 \le s \le s_0 = \pi R/2$ has two undetermined constants: $C_1$ and $\xi_0$. In order to determine them we shall use the two conditions from (50).

For the upper quarter of the corrugation wave $\varphi(s) = \int_0^s \rho(t)dt = -s/R$.

It follows from (24) and (52.1) that

$$\xi(\pi R/2) = C_1 k_{\sin}(\pi R/2) + \xi_0 = 0, \quad (53)$$

Then (36) and (52.2) together with $\sin\varphi(\pi R/2) = \sin(-\pi/2) = -1$, $\cos\varphi(\pi R/2) = 0$ yield

$$w(\pi R/2) = C_1 v_0(\pi R/2) - \xi_0 v_2(\pi R/2) = 1/2. \quad (54)$$

This way we have two equations (53) and (54) in order to determine the constants $C_1$ and $\xi_0$. In order to solve (53) and (54) we need the values $k_{\sin}(1/2)$, $v_0(1/2)$, and $v_2(1/2)$. Since $\varphi(s) = -s/R$, then (25) provides

$$k_{\sin}(s) = \int_0^s \sin\varphi(t)dt = -\int_0^s \sin(t/R)dt = R(\cos s/R - 1), \quad k_{\sin}(\pi R/2) = -R, \tag{55}$$

$$K_{\sin}(s) = \int_0^s k_{\sin}(t)dt = R\int_0^s (\cos t/R - 1)dt = R^2 \sin \pi s - Rs.$$

and, from (32), we get

$$v_0(s) = \int_0^s [H\cos^2(-t/R) - K_{\sin}(t)\sin(-t/R)]dt = \tag{56}$$

$$= \frac{1}{4}(2s(H+R^2) + R(H+R^2)\sin(\frac{2s}{R}) - 4R^3\sin(\frac{s}{R}) + 4R^2 s\cos(\frac{s}{R})),$$

$$v_2(s) = \int_0^s t\sin(-s/R)dt = R(s\cos(s/R) - R\sin(s/R)),$$

$$v_0(\pi R/2) = \frac{1}{4}(\pi HR + (\pi - 4)R^3), \quad v_2(\pi R/2) = -R^2.$$

Then, subsequently, we obtain from (53), (55), and (56) that

$$-RC_1 + \xi_0 = 0. \tag{57}$$

From (54) and (31) that

$$C_1 \frac{1}{4}(\pi HR + (\pi - 4)R^3) + \xi_0 R^2 = 1/2. \tag{58}$$

And, finally, from (57) and (58) we get

$$C_1 = \frac{2}{\pi R(H+R^2)}, \quad \xi_0 = \frac{2}{\pi(H+R^2)}. \tag{59}$$

Now we shall consider the functions concerned with the energy integral $J(x,\xi)$ in (20). The function $\varphi(s)$ has the form $\varphi(s) = -\pi s$, and then (23) and (59) yield

$$x(s) = C_1 H \cos\varphi(s) = \frac{2H}{\pi R(H+R^2)} \cos(-s/R),$$

while (24) and (59) provide

$$\xi(s) = C_1 k_{\sin}(s) + \xi_0 = \frac{2}{\pi R(H+R^2)} \cdot R(\cos \pi s - 1) + \frac{2}{\pi(H+R^2)} = \frac{2}{\pi(H+R^2)} \cos s/R.$$

The elastic energy of the half-wave equals

$$J(x,\xi) = [\frac{1}{2}\frac{Eh}{1-v^2}(\frac{2H}{\pi R(H+R^2)})^2 + \frac{1}{2}\frac{Eh^3}{12(1-v^2)}(\frac{2}{\pi(H+R^2)})^2]\int_0^{1/2} \cos^2 \pi s\, ds =$$

$$= \frac{1}{2}[\frac{Eh}{1-v^2}(\frac{2H}{\pi R(H+R^2)})^2 + \frac{Eh^3}{12(1-v^2)}(\frac{2}{\pi(H+R^2)})^2]\frac{\pi R}{4}.$$

In our example the length of the half-wave's projection on the $Ox_2$ axis is $P = R$, while for the displacement it holds that $w(1/2) = 1/2$. The corresponding macroscopic tension along the $Ox_2$ axis, i.e. the macroscopic deformation, equals $\frac{w(1/2)}{P} = \frac{1}{2R}$. Corresponding energy is $\frac{1}{2}D^0_{2222}(\frac{1}{2R})^2 P$. Then for the given corrugations profile its effective tensile stiffness $D^0_{2222}$ along

$Ox_2$ can be determined by equating the energy corresponding to the macroscopic deformation to the energy of the half-wave:

$$\frac{1}{2} D^0_{2222}(\frac{1}{2R})^2 P = \frac{1}{2}[\frac{Eh}{1-v^2}(\frac{2H}{\pi R(H+R^2)})^2 + \frac{Eh^3}{12(1-v^2)}(\frac{2}{\pi(H+R^2)})^2]\frac{\pi R}{4}.$$

Since $P = R$ and $H = h^2/12$ we obtain

$$D^0_{2222} = \frac{Eh^3}{12(1-v^2)}(\frac{2}{h^2/12+R^2})^2[\frac{h^2}{6R^2}+1]\frac{R^2}{\pi}. \tag{60}$$

For thin shells we have that $h \ll 1$ and $h^2 \ll R^2$. By dropping $h^2$, we obtain from (60) an approximate formula

$$D^0_{2222} \approx \frac{Eh^3}{12(1-v^2)}\frac{4}{\pi R^2} = \mathcal{D}\frac{4}{\pi R^2},$$

where $\mathcal{D} = \dfrac{Eh^3}{12(1-v^2)}\dfrac{4}{\pi R^2}$ is the bending stiffness of the corrugated plate.

A formula for $D^0_{2222}$ given in [5] is $\mathcal{D}\dfrac{4}{\pi(h^2/12+R^2)}$, and the one from [23], is $\mathcal{D}\dfrac{4}{\pi R^2}$ for any $h$. Once $h \ll 1$, the leading terms of both of them equal exactly $\mathcal{D}\dfrac{4}{\pi R^2}$. Also, [5] provides numerical experiments by using planar (two-dimensional) finite elements of type SHELL63 in ANSYS which confirm the above formulas for the tensile stiffness in our example.

**3. Computing the effective bending stiffness $D^2_{2222}$.** In this case, the periodicity cell problem consists of the standard equilibrium equations (14). The boundary conditions (15.5), (15.6) on the right end of the half-wave are as follows, c.f. Figs 3, 4:

$$x(L) = 0 \leftrightarrow N(L) = 0, \tag{61.1}$$
$$y(L) = 1. \tag{61.2}$$

The condition (61.1) means that while bending there is no axial force in the homogenized (i.e. two-dimensional) plate.

In this case we can solve the problem (14), (16) explicitly. By using the fact that $\varphi(L) = 0$, and also (23) together with the remarks after formula (38), the conditions (39) take the form

$$C_1 H = 0, \tag{62.1}$$
$$C_1 K^+_{\sin}(L) + \xi^+_0 L + y^+_0 = 1. \tag{62.2}$$

Then (62.1) yields $C_1 = 0$, and (62.2) provides

$$\xi^+_0 = \frac{1}{L}(1 - y^+_0). \tag{63}$$

The continuity condition for $\xi$, $y$ at $s = s_0$ has the form (40), from where we get

$$y^+_0 = C_1 K_{\sin}(\varphi_0) - s_0 C_1 k_{\sin}(\varphi_0), \tag{64}$$
$$\xi_0 = \xi^+_0 - C_1 k_{\sin}(\varphi_0).$$

Since $C_1 = 0$, then (64) yields

$$y^+_0 = 0, \ \xi_0 = \xi^+_0. \tag{65}$$

Then from (63) and (65) we obtain

$$y^+_0 = 0, \ \xi_0 = \xi^+_0 = \frac{1}{L}, \tag{66}$$

which, together with $C_1 = 0$, determine the following functions from the problem (14), (16):

$$x(s) = 0, \quad \eta(s) = 0, \tag{67}$$

$$\xi(s) = \frac{1}{L}, \quad y(s) = \frac{s}{L}, \quad 0 \leq s \leq 1. \tag{68}$$

The axial displacement and bending are a solution to the problem

$$\begin{cases} u' = w\rho(s), \\ w' = -u\rho(s) + \dfrac{s}{L}, \\ u(0) = 0, \quad w(0) = 0. \end{cases} \tag{69}$$

The problem (69) can be explicitly solved, however we do not need its solution in order to determine the effective bending stiffness. In this case we need only the energy integral $J(x,\xi)$, for which the functions we found in (67), (68) and (20) already suffice:

$$J(x,\xi) = \frac{1}{2}\frac{Eh^3}{12(1-v^2)}\int_0^L \xi_0^2 ds = \frac{1}{2}\frac{Eh^3}{12(1-v^2)}\cdot\frac{1}{L^2}L = \frac{1}{2}\frac{Eh^3}{12(1-v^2)}\frac{1}{L}. \tag{70}$$

The condition (61.2) means that the normal vector makes a 'unit rotation angle". We shall compute the macroscopic curvature corresponding to a unit the rotation angle of the normal to the homogenized (one-dimensional) plate with length $P$. In our subsequent computations in this section $x$ and $y$ are Cartesian coordinates. For a cylindrical bending we have $y'' = \rho_{cp}$ and $y = \frac{1}{2}\rho_{cp}x^2$. The unit the rotation angle of the normal vector at length $P$ means $y'(P) = 1$. Thus, we obtain $\rho_{cp}P = 1$ and the average curvature of the corrugation profile equals $\rho_{cp} = \frac{1}{P}$. Its energy is $\frac{1}{2}D_{2222}^2\left(\frac{1}{P}\right)^2 P = \frac{1}{2}D_{2222}^2\frac{1}{P}$.

The homogenized bending stiffness $D_{2222}^2$ of the corrugated plate can be expressed from the equalities of the energies of the homogenized plate and the corrugation: $\frac{1}{2}D_{2222}^2\frac{1}{P} = J(x,\xi)$. By substituting $J(x,\xi)$ from (70), we obtain

$$D_{2222}^2 = \frac{Eh^3}{12(1-v^2)}\cdot\frac{P}{L}. \tag{71}$$

The quantity $\dfrac{P}{L} \leq 1$ is (71) the ration of the corrugation wave period to its length and does not depend on the exact form of the corrugation profile.

The above formula (71) for $D_{2222}^2$ already appears in the pioneering works [1, 24]. The homogenization method confirms once again its validity.

**On the dependence of effective stiffnesses on the coordinate system.** The work [6] uses the common and widely recognized formula $\dfrac{Eh^3}{6(1-v^2)T}$ in order to compute $D_{2222}^2$ (where $T$ is the half-wave amplitude as shown in Fig.3), as well as a number of other formulas, which are distinct from each other. In this regard, we need to discuss the problem of dependence of the bending stiffnesses on our choice of the plane $x_3 = 0$ transversal to the plate: the mentioned discrepancies, although not all of them, can be explained by such dependence.

When we apply a coordinate change
$$x_\alpha \to x_\alpha \ (\alpha=1, 2), \ x_3 \to x_3 + \delta \tag{72}$$

The solution $\mathbf{N}^{1\alpha\alpha}(h,\mathbf{x})$ to problem (1) in new variables is related to its solution $\mathbf{N}^{1\alpha\alpha}(\mathbf{x})$ in old variables by $\mathbf{N}^{1\alpha\alpha}(\delta,\mathbf{x}) = \mathbf{N}^{1\alpha\alpha}(\mathbf{x}) + \delta \mathbf{X}^{0\alpha\alpha}(\mathbf{x})$, while the effective stiffnesses (4) under the coordinate change (72) are transformed as follows:

$$D^2_{\alpha\alpha\alpha\alpha}(\delta) = D^2_{\alpha\alpha\alpha\alpha} + 2\delta D^1_{\alpha\alpha\alpha\alpha} + \delta^2 D^0_{\alpha\alpha\alpha\alpha}, \ D^1_{\alpha\alpha\alpha\alpha}(\delta) = D^1_{\alpha\alpha\alpha\alpha} + \delta D^0_{\alpha\alpha\alpha\alpha}, \ D^0_{\alpha\alpha\alpha\alpha}(\delta) = D^0_{\alpha\alpha\alpha\alpha}.$$

The genuine "physical" effective stiffness of the plate is its invariant stiffness:

$$\mathbf{D}_{\alpha\alpha\alpha\alpha} = D^2_{\alpha\alpha\alpha\alpha}(\delta) - \frac{(D^1_{\alpha\alpha\alpha\alpha}(\delta))^2}{D^0_{\alpha\alpha\alpha\alpha}(\delta)}. \tag{73}$$

The above quantity does not depend on $\delta$ and thus remains invariant under the transformation (72). Apart from being invariant, (73) also enjoys the following properties: it equals the minimum of the non-invariant stiffness $D^2_{\alpha\alpha\alpha\alpha}(\delta)$, and the equality is reached for a $\delta$ such that $D^1_{\alpha\alpha\alpha\alpha}(\delta) = 0$, c.f. [18].

**Note**. In the case when the plate is homogeneous, the value of $\delta$, determined from the condition $D^1_{\alpha\alpha\alpha\alpha}(\delta) = 0$ determines the neutral plane of the plate given by $x_3 = \delta = -\dfrac{D^1_{\alpha\alpha\alpha\alpha}}{D^0_{\alpha\alpha\alpha\alpha}}$. The corresponding homogenized plate is also homogeneous, however it does not guarantee that the quantities $\dfrac{D^1_{\alpha\alpha\alpha\alpha}}{D^0_{\alpha\alpha\alpha\alpha}}$ for $\alpha = 1$ and $\alpha = 2$ coincide.

This gives rise to the question whether and when the non-invariant stiffness $D^2_{\alpha\alpha\alpha\alpha}(\delta)$ and the invariant stiffness $\mathbf{D}_{\alpha\alpha\alpha\alpha}$ coincide. It follows from (73) that $\mathbf{D}_{\alpha\alpha\alpha\alpha} = D^2_{\alpha\alpha\alpha\alpha}(\delta)$ once $D^1_{\alpha\alpha\alpha\alpha}(\delta) = 0$.

By multiplying (3.1) by $y_2 \delta_{2\delta}$ when $\nu = 1$ and by further integrating it provided boundary conditions (3.2) and (3.3), we obtain

$$\int_{P_0} a_{22\iota\kappa} M^1_{\iota,\kappa} dx_2 dx_3 = 2P \int_{x_2=P} a_{22\iota\kappa} M^1_{\iota,\kappa}(P,x_3) dx_3 = 2P \int_{y_3=P} \sigma_{22}(P,x_3) dy_3 \tag{74}$$

for a corrugation profile that is symmetric with respect to its medial plane.

In (74) we use the fact that the strains on the right $x_2 = P$ and left $x_2 = -P$ boundaries of the periodicity cell coincide, while the corresponding normals are opposite, c.f. Fig.2). The corresponding displacements equal $P$ and $-P$, as in (3.3). Then, from (4) and (74), we obtain

$$D^1_{2222} = \frac{1}{2P} \int_{P_0} a_{22\iota\kappa} M^1_{\iota,\kappa}(x_2,x_3) dx_2 dx_3 = \int_{y_3=P} \sigma_{22}(P,x_3) dx_3 \tag{75}$$

where stress $\sigma_{22} = a_{22\iota\kappa} M^1_{\iota,\kappa}$.

In the case of a thin plate, the integral in (75) is the normal force on the right end of the corrugation profile $N(L)$. Thus, the condition that $N(L) = 0$ implies $D^1_{2222} = 0$. The above solution is obtained under the assumption (61.1) $N(L) = 0$. Then the stiffness computed in (71) equals the invariant stiffness $\mathbf{D}_{2222}$.

**4. Computing the effective stiffnesses $D^0_{1122}$, $D^2_{1122}$, $D^0_{1111}$, and $D^2_{1111}$. The universal relations between stiffnesses.** Relations between homogenized constants are called universal if they do not depend on the specific microscopic structure of the solid [25].

It seems that the easiest way to compute the above-mentioned quantities is by using the universal relations between the corrugated plate's homogenized stiffnesses established in [8]. Namely, the relations from [8] take the following form in our present notation:

$$D^0_{2211} = \nu D^0_{2222}, \quad D^2_{2211} = \nu D^2_{2222} \tag{76}$$

and

$$D^0_{1111} = \nu^2 D^0_{2222} + E\frac{1}{2P}\int_{P_0} dx_2 dx_3, \quad D^2_{1111} = \nu^2 D^2_{2222} + E\frac{1}{2P}\int_P x_3^2 dx_2 dx_3. \tag{77}$$

Let us compute the integral in (77). We obtain ($n$ means coordinate along the normal vector **n**)

$$\int_P dxdy = 2\int_0^L ds \int_{-h/2}^{h/2} dn = 2hL,$$

$$\int_{P_0} x_3^2 dx_2 dy_3 = 2\int_0^L \int_{-h/2}^{h/2} [x_3(s) + n\cos\varphi(s)]^2 dn\, ds = 2h\int_0^L x_3(s)^2 ds + 2\frac{h^3}{12}\int_0^L \cos^2\varphi(s)ds.$$

Then

$$D^0_{1111} = \nu^2 D^0_{2222} + \frac{EhL}{P}, \quad D^2_{1111} = \nu^2 D^2_{2222} + \frac{E}{P}[h\int_0^L x_3(s)^2 ds + \frac{h^3}{12}\int_0^L \cos^2\varphi(s)ds]. \tag{78}$$

The formulas (78) above have a good deal of proximity to [6, formulas (13) and (27)]. Namely, they coincide if we do not expand the terms $D^0_{2222}$ and $D^2_{2222}$ of the latter. However, our expression for $D^0_{2222}$ and $D^2_{2222}$ are different from those in [6]. The only reason for this discrepancy is the choice of coordinate system, as discussed above. The first term in the brackets in formula (78) has order 1 in $h$, while the other two have order 3 in $h$. Thus, if $h \ll 1$ we can rewrite (78) as an approximate formula

$$D^2_{1111} \approx \frac{Eh}{P}\int_0^L x_3(s)^2 ds. \tag{79}$$

**5. The effective stiffnesses** $D^{\nu+\mu}_{1212}$, $D^{\nu+\mu}_{2121}$. The periodicity cell problems of homogenization theory for the mentioned stiffnesses are reduced to a single differential equation that corresponds to antiplanar deformation, c.f. [8, 15]. Namely, if $\nu = 0$ then the periodicity cell problem takes the form of a Laplace equation

$$\begin{cases} \Delta M = 0 \text{ in } P_0, \\ \frac{\partial M}{\partial \mathbf{n}} = 0 \text{ on } \Gamma_0, \\ M(\pm L, y_3) = \pm L. \end{cases} \tag{80}$$

Then the formula for the effective stiffness becomes

$$D^0_{1212} = \frac{1}{2P}\int_{P_0} a_{1212}(x_2, x_3) M_{,2}(x_2, x_3) dx_2 dx_3. \tag{81}$$

If $\nu = 1$, then the periodicity cell problem takes the form of a Poisson equation, c.f. [8]:

$$\begin{cases} \Delta \psi = -1 \text{ in } P_0, \\ \psi(x_2, x_3) = 0 \text{ on } \Gamma_0, \\ \psi(x_2, x_3) \text{ periodic in } x_2 \in [-L, L] \end{cases} \tag{82}$$

and the respective formula for the effective stiffness is given by [8]

$$D_{1212}^2 = -\frac{1}{2P}\int_{P_0} a_{1212}\psi_{,3}(x_2,x_3)x_3 dx_2 dx_3 . \tag{83}$$

The problems (80), (82) can be treated as thermal conductivity problems since the latter are well-understood [26].

**Solving the problem (80)**. It is more convenient to solve (80), as well as (82), on the whole corrugation wave, and not on the half-wave.

The natural coordinate system $\{\boldsymbol{\tau},\mathbf{n}\}$ with the respective coordinate ($s$, $n$) is orthonormal. By [26], the Laplacian is invariant under a coordinate change from one orthonormal system to another. Then (80) takes the following form in the natural coordinates ($s$, $n$):

$$\begin{cases} \dfrac{\partial^2 M}{\partial s^2} + \dfrac{\partial^2 M}{\partial n^2} = 0 \text{ in } P_0, \\ \dfrac{\partial M}{\partial y} = 0 \text{ for } y = \pm h/2, \\ M(\pm L, n) = \pm L. \end{cases} \tag{84}$$

We can verify by direct substitution that (84) has the following solution:
$$M(s, y) = s . \tag{85}$$

Further, $\partial/\partial y_2 = \cos\varphi(s)\partial/\partial s + \sin\varphi(s)\partial/\partial n$ and for the function in (85) we have $\partial M/\partial y_2 = \cos\varphi(s)$. Thus the integral in (81) becomes

$$\int_{-L}^{L}\int_{-h/2}^{h/2} \cos\varphi(s)dsdn = h\int_{-L}^{L}\cos\varphi(s)ds .$$

Finally,

$$D_{1212}^0 = \frac{1}{2P}\int_{P_0} a_{1212}(x_2,x_3)M_{,2}(x_2,x_3)dx_2 dx_3 = a_{1212}\frac{h}{P}\int_0^L \cos\varphi(s)ds = a_{1212}h . \tag{86}$$

We compute $\int_0^L \cos\varphi(s)ds = \int_0^L x_2'(s)ds = x_2(L) - x_2(0) = P$, and since $a_{1212} = \dfrac{E}{2(1+\nu)}$, formula (86) provides

$$D_{1212}^0 = \frac{Eh}{2(1+\nu)} . \tag{87}$$

There are different opinions in the literature concerning the computation of $D_{1212}^0$. E.g. [6] mentions $D_{11} = \dfrac{P}{L}\cdot\dfrac{Eh}{2(1+\nu)}$ (in our present notation) as a standard and widely accepted formula (c.f. formulas (15) and (17) in [6]). Another formula given in [27] (c.f. formulas (16) and (17) in [6]) coincides in (87). The homogenization method verifies the ones in [27].

**Solving the problem (82)**. The problem (82) takes the following form in the natural coordinates:

$$\begin{cases} \dfrac{\partial^2 \psi}{\partial s^2} + \dfrac{\partial^2 \psi}{\partial n^2} = -1 \text{ in } P_0 \\ \psi(s,\pm h/2) = 0 \text{ on } \Gamma_0, \\ \psi(s,n) \text{ periodic in } s\in[-L,L]. \end{cases} \tag{88}$$

Its solution is

$$\psi(s,n) = -\frac{1}{2}(n+h/2)(n-h/2) = -\frac{1}{2}(n^2 - \frac{h^2}{4}). \tag{89}$$

This can be verified by substituting (89) into (88) directly. Moreover,

$$y_3 = y_3(s) + n\cos\varphi(s), \quad \frac{\partial}{\partial y_3} = \frac{\partial}{\partial n}\cos\varphi(s) + \frac{\partial}{\partial s}\sin\varphi(s)$$

and for the function in (89) we have $\frac{\partial \psi}{\partial y_3} = -n\cos\varphi(s)$. Thus, the integral in (83) becomes

$$\int_{-L}^{L}\int_{-h/2}^{h/2} n\cos\varphi(s)(y_3(s) + n\cos\varphi(s))dsdn = \frac{h^3}{12}\int_{-L}^{L}\cos^2\varphi(s)ds.$$

In the formula above we use the fact that $\int_{-h/2}^{h/2} ndn = 0$. Then, finally, we arrive at

$$D_{1212}^2 = \frac{Eh^3}{24(1+\nu)P}\int_{-L}^{L}\cos^2\varphi(s)ds. \tag{90}$$

Here we have to stress the fact that computing $D_{1212}^2$ by the classical homogenization method is not practically feasible for thin plates because of the high numerical instability of its formulas for homogenized stiffnesses. Namely, the formula for $D_{1212}^2$ implies computing a difference of two integrals whose values are much bigger than the said difference [8]. This makes the stiffness $D_{1212}^2$ sharply distinct from other stiffnesses, for which the homogenization method works correctly. Because of this issue, the problem (82) for computing the value of $D_{1212}^2$ was introduced in [8], and its solution provided numerically stable formulas. For thin plates, (82) provides a simple formula for $D_{1212}^2$ given in (90).

**Table 1.** Formulas for effective stiffnesses of corrugated plates

| | $D_{2222}^0$ | $D_{2222}^2$ | $D_{1122}^0, D_{1122}^2$ | $D_{1111}^0, D_{1111}^2$ | $D_{1212}^0$ | $D_{1212}^2$ |
|---|---|---|---|---|---|---|
| C.f. (50) | | $\frac{Eh^3}{12(1-\nu^2)} \cdot \frac{P}{L}$ | $D_{2211}^0 = \nu D_{2222}^0$, $D_{2211}^2 = \nu D_{2222}^2$ | $\nu^2 D_{2222}^0 + \frac{EhL}{P}$, $\approx \frac{Eh}{P}\int_0^L x_3(s)^2 ds$ | $\frac{Eh}{2(1+\nu)}$ | $\frac{Eh^3}{24(1+\nu)P}\int_{-L}^{L}\cos^2\varphi(s)ds$ |

In Table 1, we collected all the effective stiffnesses for corrugated plates, and the corresponding formulas. Only $D_{2222}^2$ and $D_{1212}^0$ have explicit formulas, while $D_{1122}^0$ and $D_{1122}^2$ are explicitly related to other stiffnesses from the table. The values of $D_{2222}^2$ and $D_{1212}^0$ do not depend on the actual form of the corrugation profile. The formulas (78) and (90) for $D_{1111}^2$, $D_{1212}^2$ are simple enough, although not explicit. In-plane stifnesses $D_{1111}^0$ and $D_{1122}^0$ are expressed through $D_{2222}^0$. The computing the tensile effective stiffness $D_{2222}^0$ in the plane transversal to the corrugation wave requires some effort. It is unlikely that such computations can be performed symbolically. However, all the formulas from Table 1 can easily be used for numerical computations. The above procedure for computing $D_{2222}^0$, can be realized as a computer routine.

**Conclusions**. In the present paper we compute the effective stiffnesses for corrugated plates by approximating the corresponding periodicity cell problem with a homogenized problem for curvilinear beams. Such an approach is the closest one to homogenization theory, while we solve a periodicity problem of homogenization theory by methods specific to elasticity theory.

The resulting effective stiffnesses turn out to be dependent only on a single function, which is the curvature of the corrugation profile (all the geometric parameters in (79) and (90), such as the angle and the amplitude of the corrugation wave can be expressed through its curvature).

The computation of effective stiffnesses, except in-plane stiffnesses $D^0_{1111}$, $D^0_{1122}$ and $D^0_{2222}$, is performed either by explicit formulas, or by calculating the values of relatively simple integrals related to the angle and curvature of the corrugation wave. In some cases, these integrals can be computed in explicit form, and they can be always computed numerically. The computation of effective in-plane stiffnesses $D^0_{1111}$, $D^0_{1122}$ and $D^0_{2222}$ is reduced to the computation of $D^0_{2222}$. This requires some effort. We note that the procedure for computation of $D^0_{2222}$ demonstrates that there exists no simple exact formula for computation of $D^0_{2222}$ (thus, $D^0_{1111}$, $D^0_{1122}$) for corrugation of arbitraru shape.

We have considered in detail only smooth corrugation profiles (i.e. those without creases and flats) which split into two intervals where the curvature of the corrugation wave keeps constant sign. Our method is generalizable to more than two such intervals, which requires merging solutions in more than a single point.

As well, our method can be adapted to corrugation profiles with creases and flats. Namely, in the cases of creases we need to address the issue of merging solutions at each such a crease. This is doable, since the problem of joining beams at non-smooth angles is solved in [28, 29]. As for the case of flats, at each flat we deal with a linear beam, and the problem actually simplifies.

However, we need to merge all the solutions that we obtained on different intervals, which can be hard to do manually provided that the number of intervals can be great. It seems that this issue can be reasonably addressed by developing a computer routine.

**References**

1. Huber, M.T. (1923) Die Theorie des kreuzweise bewehrten Eisenbetonplatten. *Der Bauingenieur*, 4, 354–360.
2. Buannic, N., Cartraud, P., Quesnel, T. (2003) Homogenization of corrugated core sandwich panels. *Composite Struct*, 59, 299-312.
3. Talbi, N., Batti, A., Ayad, R., Guo, Y.Q. (2009) An analytical homogenization model for finite element modelling of corrugated cardboard. *Composite Struct*, 88, 280–289.
4. Lee, C.Y., Yu, W. (2011) Homogenization and dimensional reduction of composite plates with inplane heterogeneity. *Int J Solids Struct*, 2011, 48, 1474–1484.
5. Xia, Y., Friswell, M.I., Flores, E.I.S. (2012) Equivalent models of corrugated panels. *Int J Solids Struct*, 49(20), 1453–1462.
6. Ye, Z., Berdichevsky, V.L., Yu, W. (2014) An equivalent classical plate model of corrugated structures. *Int J Solids Struct*, 51(11-12), 2073–2083
7. Bartolozzi, G., Baldanzini, N., Pierini, M. (2014) Equivalent properties for corrugated cores of sandwich panels: a general analytical method. *Composite Struct*, 108, 736–746.
8. Annin, B.D., Kolpakov, A.G., Rakin, S.I. (2017) Homogenization of corrugated plates based on the dimension reduction for the periodicity cell. *Problem Mechanics for Materials and Technologies*. H. Altenbach et al. (eds.), Springer Inter. Publ. AG., 30-72.
9. Aoki, Y., Maysenhölder, W. (2017) Experimental and numerical assessment of the equivalent-orthotropic-thin-plate model for bending of corrugated panels. *Int J Solids Struct*, 108 11–23

10. Dayyani, I., Shaw, A.D., Saavedra Flores, E.I., Friswell, M.I. (2015) The mechanics of composite corrugated structures: A review with applications in morphing aircraft. *Composite Struct*, 133, 358-380
11. Caillerie, D. (1984) Thin elastic and periodic plates. *Math Methods Appl Sciences*, 6, 159–191.
12. Kohn, R.V., Vogelius, M. (1984) A new model for thin plates with rapidly varying thickness. *Int J Solids Struct*, 20, 333–350.
13. Winkler, M.Ch. (2012) *Analysis of Corrugated Laminates*. A dissertation submitted to ETH Zurich for the degree of Doctor of Sciences Diss. ETH No. 20130.
14. Andrianov, I.V., Comas-Cardona, D. Binetruy, C. (2013) Corrugated beams mechanical behavior modeling by the homogenization method. *Int J Solids Struct,* 50 928–936
15. Kolpakov, A.G., Rakin, S.I. (2016) Calculation of the effective stiffnesses of corrugated plates by solving the problem on the plate cross-section. *J Appl Mech Tech Phys*, 57(4), 757-767.
16. Kolpakov, A.G. (1985) Determination of the average characteristics of elastic frameworks. *J Appl Math Mech*. 49 (6), 739–745.
17. Pontryagin, L. S., Boltyanskii, V. G., Gamkrelidze, R. V., Mishechenko, E. F. (1962) *The Mathematical Theory of Optimal Processes*. NY: John Wiley & Sons.
18. Kolpakov, A.G. (2004) *Stressed Composite Structures. Homogenized Models for Thin-Walled Nonhomogeneous Structures with Initial Stresses*. Hidelberg: Springer.
19. Kalamkarov, A.L., Kolpakov, A.G. (1997) *Analysis, Design and Optimization of Composite Structures*. Chichester: John Wiley&Sons.
20. Washizu, K. (1982) *Variational Methods in Elasticity and Plasticity*. 3rd Ed., New York: Pergamon Press.
21. Ye, Z., Yu, W. (2013) Homogenization of Piecewise Straight Corrugated Plates, *54th AIAA/ASME/ASCE/AHS/ASC Structures, Structural Dynamics, and Materials Conference* (AIAA 2013-1608) https://arc.aiaa.org/doi/abs/10.2514/6.2013-1608.
22. Arnol'd, V. I. (1973) *Ordinary differential equations*, Cambridge, MA: MIT Press.
23. Yokozeki, T., Takeda, S.-I., Ogasawara, T., Ishikawa, T., (2006). Mechanical properties of corrugated composites for candidate materials of flexible wing structures. *Composites: Part A*, 37, 1578–1586.
24. Seydel, E. (1941) Shear buckling of corrugated plates. *Jahrbuch die Deutschen Versuchsanstalt fur Luftfahrt*, 9, 233-245.
25. Dvorak, G. (2013) *Micromechanics of Composite Materials*. Berlin: Springer.
26. Carslaw, H.S., Jaeger, J.C. (1959) *Conduction of Heat in Solids*, 2nd Ed., Oxford: Oxford University Press.
27. Briassoulis, D. (1986) Equivalent orthotropic properties of corrugated sheets. *Comput. Struct*. 23, 129–138.
28. Zhikov, V.V. (2002) Homogenization of elasticity problems on singular structures. *Izvestia: Mathematics*, 66 (5), 299–365.
29. Zhikov, V.V., Pastukhova, S.E. (2003) Homogenization for elasticity problems on periodic networks of critical thickness. *Sbornik: Mathematics*, 194 (8), 697–732.